\documentclass{emulateapj}
\usepackage{epsfig}
\usepackage{color}

\newcommand   {\about} {\mbox{$\sim$}}

\newcommand   {\arcs}  {\mbox{$^{\prime\prime}$}}
\newcommand   {\ps}    {\mbox{s$^{-1}$}}

\newcommand   {\kms}   {\mbox{km\,s$^{-1}$}}
\renewcommand {\ga}    {\mbox{\rlap{\hbox{\lower4pt\hbox{$\sim$}}}\hbox{$>$}}}
\renewcommand {\la}    {\mbox{\rlap{\hbox{\lower4pt\hbox{$\sim$}}}\hbox{$<$}}}

\received{4 May 2007}
\accepted{27 November 2007}

\slugcomment{To appear in The Astrophysical Journal}

\shorttitle{HNC in Comets}
\shortauthors{Lis et~al.}

\begin{document}

\title{Hydrogen Isocyanide in Comet 73P/Schwassmann-Wachmann (Fragment
B)}

\author{D.C.~Lis\altaffilmark{1,4},
D.~Bockel\'{e}e-Morvan\altaffilmark{2}, J.~Boissier\altaffilmark{2},
J.~Crovisier\altaffilmark{2}, N.~Biver\altaffilmark{2}, and
S.B.~Charnley\altaffilmark{3}
}

\altaffiltext{1}{California Institute of Technology, Downs Laboratory
of Physics 320-47, Pasadena, CA~91125; dcl@caltech.edu}

\altaffiltext{2}{LESIA and UMR 8109 du CNRS, Observatoire de Paris, 5
place Jules Janssen, 92195 Meudon, France;
Dominique.Bockelee@obspm.fr, Jeremie.Boissier@obspm.fr,
Jacques.Crovisier@obspm.fr; Nicolas.Biver@obspm.fr}

\altaffiltext{3}{Space Science and Astrobiology Division, MS 245--3,
  NASA Ames Research Center, Moffet Field, CA 94035;
  charnley@dusty.arc.nasa.gov}

\altaffiltext{4}{Visiting Scientist, LESIA, Observatoire de Paris}

\begin{abstract}
We present a sensitive 3-$\sigma$ upper limit of 1.1\% for the HNC/HCN
abundance ratio in comet 73P/Schwassmann-Wachmann (Fragment B),
obtained on May 10--11, 2006 using Caltech Submillimeter Observatory
(CSO). This limit is a factor of \about 7 lower than the values
measured previously in moderately active comets at 1~AU from the Sun.
Comet 73P/Schwassmann-Wachmann was depleted in most volatile species,
except of HCN. The low HNC/HCN ratio thus argues against HNC
production from polymers produced from HCN. However, thermal
degradation of macromolecules, or polymers, produced from ammonia and
carbon compounds, such as acetylene, methane, or ethane appears a
plausible explanation for the observed variations of the HNC/HCN ratio
in moderately active comets, including the very low ratio in comet
73P/Schwassmann-Wachmann reported here. Similar polymers have been
invoked previously to explain anomalous $^{14}$N/$^{15}$N ratios
measured in cometary CN.
\end{abstract}


\section{Introduction}

Hydrogen isocyanide, HNC, a metastable isomer of HCN, was first
detected in the ISM over a third of a century ago as the unidentified
line U90.7 (\citealt{snyder72, zucker72}; the identification
subsequently confirmed in the laboratory by \citealt{black76}). The
interstellar HNC/HCN abundance ratio has been shown to be strongly
temperature dependent (e.g., \citealt{schilke92}; \citealt{hirota98},
and references therein). The current understanding of the HNC
chemistry in molecular hot cores is that in warm, dense gas HCN is
first produced efficiently by ion-molecule chemistry from ammonia and
C$^+$. HNC is subsequently produced by proton transfer to HCN to form
HCNH$^+$, followed by dissociative recombination or proton transfer to
ammonia (e.g., \citealt{rodgers01a, charnley02}). The temperature
dependence of the HNC/HCN abundance ratio is explained primarily by
proton transfer reactions cycling between the two isomers via the
HCNH$^+$ ion. However, additional HNC formation routes may be required
in dark clouds where the HNC/HCN ratio exceeds unity. These are
presumably neutral-neutral reactions (e.g., C+NH$_2$=HNC+H;
\citealt{herbst00}) and consequently are probably too slow to
efficiently form HNC in expanding cometary atmospheres.

The initial detection of HNC in a cometary atmosphere was by
\cite{irvine96} in comet C/1996 B2 (Hyakutake). The measured HNC/HCN
abundance ratio, \about 6\%, was similar to that in interstellar
clouds with a temperature of order 50~K, suggesting that cometary HNC
may be unprocessed interstellar material incorporated into the comet's
nucleus. However, \cite{irvine96} argued that a number of alternative
processes may also explain the observed HNC/HCN ratio in comet
Hyakutake, including irradiation of icy matrix containing HCN,
non-equilibrium chemical processes in the solar nebula, gas-phase
processes in the coma itself, infrared relaxation of HCN from excited
vibrational levels of the ground electronic state, or
photo-dissociation of a heavier parent molecule.

A strong variation of the HNC/HCN abundance ratio in comet C/1995 O1
(Hale-Bopp) with heliocentric distance (from \la 2\% at 2.9~AU to
\about 20\% near 1~AU; \citealt{biver97}; \citealt{irvine98})
questioned the interstellar origin of cometary HNC and suggested a
production mechanism in the coma itself as a more likely explanation.
\cite{rodgers98} presented a comprehensive model of the cometary coma
chemistry and suggested that in very active comets, such as comet
Hale-Bopp, the observed variation of the HNC abundance with the
heliocentric distance, as observed with single-dish telescopes, can be
explained by isomerization of HCN driven by the impact of fast
hydrogen atoms produced in photo-dissociation of parent molecules.
However, their model overproduced the HNC abundance at \about 3~AU by
about a factor of 2. In a subsequent paper \cite{rodgers01b} showed
that the same mechanism cannot reproduce observed HNC/HCN abundance
ratios in moderately active comets at \about 1~AU, such as comet
C/1999 H1 (Lee). The applicability of the model to very active comets
has also been questioned by interferometric observations of HNC and
HCN in comet Hale-Bopp \citep{bockel05, bockel07}, which show that the
HNC/HCN visibility ratio is almost flat as a function of baseline, and
equal to the single-dish flux ratio. This suggests that, contrary to
model predictions, HNC in comet Hale-Bopp was produced efficiently in
the inner coma (parent scale-length $\ll 2000$~km).

Comet 73P/Schwassmann-Wachmann is a short-period (5.34 years) comet of
the Jupiter family that likely originated from the trans-Neptunian
population in the outer Solar System. Narrow-band photometry and
spectroscopic observations during the 1990 apparition showed that this
comet is strongly depleted in C$_2$ and NH$_2$ relative to CN and OH,
placing it in the 21P/Giacobini-Zinner class of odd comets with
extreme depletions \citep{ahearn95,fink96}. During its 1995 perihelion
passage, comet 73P split into several fragments \citep{boeh02}. Its
2006 apparition thus offered an exceptional opportunity to measure the
HNC/HCN ratio in fragments of a Jupiter-family comet of deviant
composition. Indeed, comet 73P made a very close approach to Earth in
May 2006 (at 0.067 and 0.079~AU for Fragments B and C, respectively),
that compensated for its moderate activity (water production rate of a
few $\times$ 10$^{28}$ s$^{-1}$ at perihelion) and allowing sensitive
investigations.

In the present paper, we compile the existing measurements of the
HNC/HCN ratio in moderately active comets, including our
sensitive upper limit obtained in comet 73P/Schwassmann-Wachmann and
discuss possible implications for the origin of HNC in cometary
atmospheres.

\section{Observations}

Observations of Comet 73P/Schwassmann-Wachmann (Fragment B)
presented here were carried out on May 9--11, 2006 UT using the 10.4-m
Leighton telescope of Caltech Submillimeter Observatory (CSO) on Mauna
Kea in Hawaii. The heliocentric and geocentric distances of the comet
at the time of the CSO observations were 1.02 and 0.072~AU,
respectively. We used the 345 GHz facility SIS receiver and
acousto-optical spectrometers with total bandwidths of 50 MHz and 1
GHz. The frequency scale of the spectrometers (the reference channel
and the channel width) was established by 
injecting calibration signals from 10~MHz and 100~MHz frequency
comb generators, for the 50~MHz and 1~GHz AOS, respectively. The
effective frequency resolution of the high-resolution (50 MHz
bandwidth) spectrometer was 95 kHz (2 channels), corresponding to
\about 0.08~\kms. Pointing of the telescope, checked by performing
five-point continuum scans of Jupiter in the south, was determined to
drift by \la 5\arcs\ on time-scales of \about 3 hours at the time of
the observations (one fifth of the \about 25\arcs\ FWHM CSO beam
at 360 GHz). The accuracy of the cometary ephemeris, checked by
performing five-point scans of the HCN~$J$=4--3 transition in the
comet, was determined to be \about 7--9\arcs. The main beam
efficiency, determined from total-power observations of Jupiter,
Saturn and Mars, was measured to be \about 74\%. The absolute
calibration uncertainty of the individual measurements is \about 20\%.
However, since the HCN and HNC lines were observed almost
simultaneously, with the same receiver, we estimate the calibration
uncertainty of the HNC/HCN line ratio to be smaller, of order 15\%.

\begin{figure}
\begin{center}
\epsfig{file=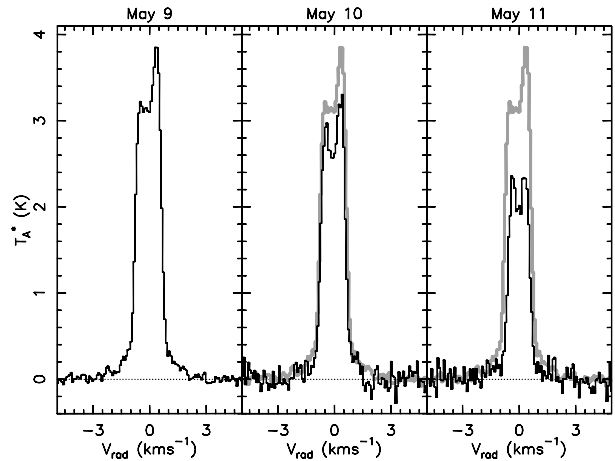,width=6cm,angle=270}
\end{center}
\caption{\label{fig_hcn} HCN~$J$=4--3 spectra in comet
  73P/Schwassmann-Wachmann (Fragment B) obtained on May 9--11 (left
  to right, respectively). Gray curves in the two right panels
  show the May 9 spectrum.}
\end{figure}

\begin{figure}
\begin{center}
\epsfig{file=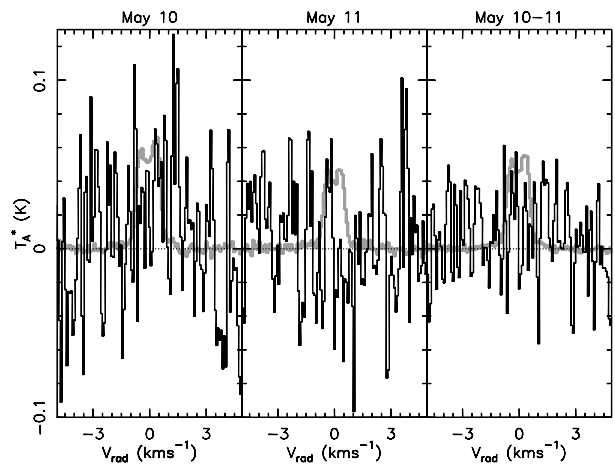,width=6cm,angle=270}
\end{center}
\caption{\label{fig_hnc} HNC~$J$=4--3 spectra in comet
  73P/Schwassmann-Wachmann obtained on May 10--11 (left and middle
  panels, respectively). The right panel shows an uniformly weighted
  average of the two data sets. Gray curves show HCN~$J$=4--3 spectra
  on the corresponding days, scaled down by a factor of 50.}
\end{figure}

HCN~$J$=4--3 spectra (rest frequency 354.50547 GHz) taken on May 9--11
UT are shown in Figure~\ref{fig_hcn} and the line intensities are
listed in Table~\ref{tab_int}, along with the corresponding
statistical uncertainties, as determined from the noise level
measured in the spectra. A significant decrease in the HCN intensity
(by a factor of 1.8) is seen over the period of our observations,
following an earlier outburst. The low-level pedestal, seen most
clearly in the May 9 spectrum, is due to the HCN hyperfine structure
(e.g., \citealt{lis97}). HNC~$J$=4--3 spectra (rest frequency
362.63031 GHz) taken on May 10--11 UT are shown in
Figure~\ref{fig_hnc}. The HNC and HCN observations on these two
nights were interleaved in time to account for the time variability of
the outgassing. No HNC emission is detected (Table~\ref{tab_int}).

\begin{table}
  \begin{center}
    \caption{HCN and HNC Line Intensities.}\vspace{1em}
    \renewcommand{\arraystretch}{1.2}
    \footnotesize
    \begin{tabular}{lcc}
      \hline
      \hline
      $UT Date$        & $I({\rm HCN})$ & $I({\rm HNC})$  \\
      \hline
      May \,~9, 2006      & $5.35 \pm 0.04$ &  ---         \\
      May 10, 2006     & $4.21 \pm 0.08$ & $~~0.073 \pm 0.025$ \\
      May 11, 2006     & $3.01 \pm 0.06$ & $-0.021 \pm 0.018$ \\
      May 10--11, 2006 & $3.62 \pm 0.05$ & $~~0.026 \pm 0.015$ \\
      \hline
     \end{tabular}
    \label{tab_int}
  \end{center}

Notes.---Integrated line intensities, $T_A^* dv$ (K\,\kms), computed
over the velocity range of --2 to 2 \kms\ (HCN; includes all hyperfine
components) and --1 to 1 \kms\ (HNC). The uncertainties listed are
1-$\sigma$ statistical uncertainties, as determined from the
noise level measured in the spectra. The May 10--11 entry is a
uniformly weighted average of the two corresponding spectra.
\end{table}

\begin{table*}[!ht]
  \begin{center}
    \caption{HCN and HNC Production Rates.}\vspace{1em}
    \renewcommand{\arraystretch}{1.2}
    \footnotesize
    \begin{tabular}{lccc}
      \hline
      \hline
      $UT Date$        & $Q({\rm HCN})$ & $Q({\rm HNC})$ & $Q({\rm
      HCN})$/$Q({\rm HNC})$  \\ 
      \hline
      May 10, 2006     & $(4.85 \pm 0.09) \times 10^{25}$ & 
      $<0.7 \times 10^{24}$  & $<1.4\%$ \\ 
      May 11, 2006     & $(3.05 \pm 0.06) \times 10^{25}$ &
      $<0.5 \times 10^{24}$  & $<1.7\%$ \\
      May 10--11, 2006 & $(3.90 \pm 0.05) \times 10^{25}$ &
      $<0.4 \times 10^{24}$ & $<1.1\%$ \\
      \hline
    \end{tabular}
    \label{tab_prod}
  \end{center}

Notes.---Productions rates (mol~s$^{-1}$) computed assuming a gas
temperature of 65~K and expansion velocities of 0.62 and 0.58 \kms,
on May 10 and 11, respectively. Average pointing offsets of 3\arcs\
and 4\arcs\ for HCN and HNC, respectively, have been assumed. Since
the HCN and HNC transitions studied here have similar rest frequencies
and line strengths, the derived Q(HNC)/Q(HCN) ratio is insensitive to
the detailed model assumptions.
\vspace{1em}
\end{table*}

Fragment B of comet 73P underwent several outbursts during the
April--May 2006 apparition, shedding tens of fragments, as imaged for
example by the Hubble Space Telescope around 18 April (e.g.,
\citealt{toth06}). On 8.0 May UT fragment B entered a new outburst
phase, beginning as a strong brightness increase of its central
condensation. Visual magnitudes reported in \cite{icq06} showed a
fivefold increase in total brightness the following day and a 5 to 10
times increase in the water outgassing rate was seen in the Nan\c{c}ay
and Odin data between 7 and 10 of May \citep{crovi06, colom06}.

Molecular production rates of HCN and HNC (3-$\sigma$ upper limits)
on May 10--11 are listed in Table 2. The resulting 3-$\sigma$ upper
limit for the HNC/HCN abundance ratio, based on the average of the May
10 and May 11 data is 1.1\%. IRAM 30-m observations impose upper
limits of 2.9 and 3.8\% for the HNC/HCN ratio on May 14 and 18,
respectively \citep{biver06a}, consistent with the low value
presented here. 

\section{HNC/HCN Ratio in Comets}

\begin{figure}
\begin{center}
\epsfig{file=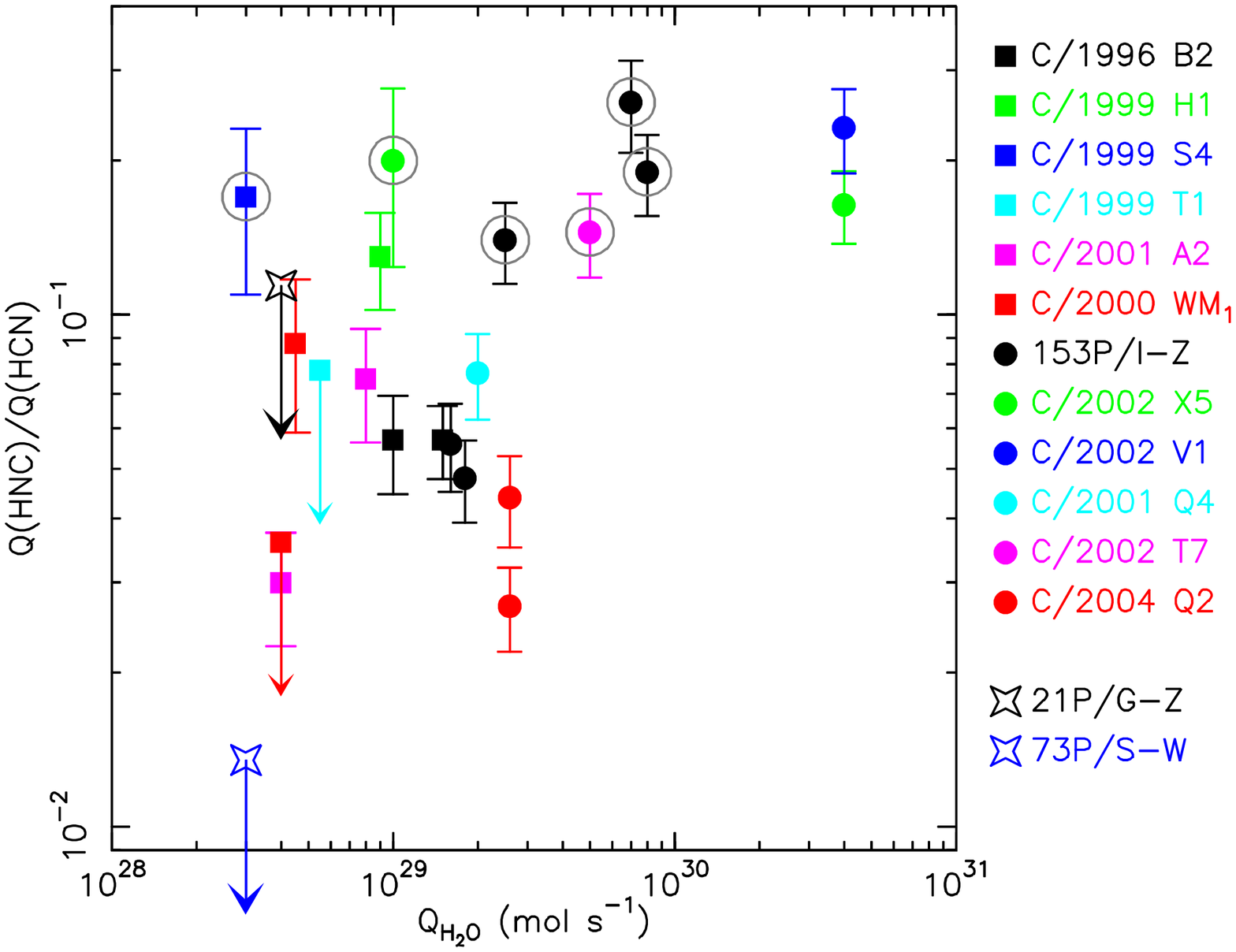,width=6cm,angle=270}
\end{center}
\caption{\label{fig_prod} Observed HNC/HCN abundance ratio in comets
  as a function of the water production rate, $Q_{\rm H_2O}$.
  Measurements at the heliocentric distances between 0.5 and 0.8~AU
  are surrounded by gray circles. References: C/1996 B2 (Hyakutake):
  \citealt{irvine96}; C/1999 H1 (Lee): \citealt{biver00}; C/1999 S4
  (LINEAR): \citealt{bockel01}; C/1999 T1 (McNaught-Hartley):
  \citealt{biver06b}; C/2000 WM$_1$ (LINEAR): \citealt{biver06b},
  \citealt{irvine03}; C/2001 A2 (LINEAR): \citealt{biver06b};
  153P/Ikeya--Zhang: \citealt{biver06b}, \citealt{irvine03}; C/2001 Q4
  (NEAT), C/2002 T7 (LINEAR), C/2002 V1 (NEAT), C/2002 X5
  (Kudo-Fujikawa), and C/2004 Q2 (Machholz): \cite{biver05},
  \citealt{crovi05}; 21P/Giacobini--Zinner: \citealt{biver99};
  73P/Schwassmann--Wachmann: this work. The combined
  uncertainties of the individual measurements in the figure are
  assumed to be equal to the statistical uncertainties, as measured
  from the individual spectra, plus a 15\% calibration uncertainty,
  added in quadrature.}
  \vspace{1em}
\end{figure}

HNC has now been detected in 11 comets with water production rates in
the range $3 \times 10^{28}- 4 \times 10^{30}$~mol\,s$^{-1}$ (e.g.,
\citealt{irvine96}; \citealt{biver05}; \citealt{crovi05};
\citealt{biver06b}), not counting the very active comets Hale-Bopp and
C/2006 P1 (McNaught), where different chemical processes may be at
work. The observed HNC/HCN abundance ratio is plotted in
Figure~\ref{fig_prod} as a function of the water production rate. The
plot appears to show a distinction between comets with water
production rates above and below \about $4 \times
10^{29}$~mol\,s$^{-1}$. More productive comets seem to display a
relatively high HNC/HCN ratio, \about 0.2, with little scatter. The
scatter increases significantly in weaker comets and no clear
dependence of the observed abundance ratio on the water production
rate is seen. Since water production rates in comets vary with the
heliocentric distance and the measurements included in our sample span
a wide range in $r_h$, we have outlined with gray circles in
Figure~\ref{fig_prod} the measurements with $r_h$ limited to a narrow
range of 0.5--0.8~AU. This sub-sample includes four comets, which
cover a wide range of water production rates. No
dependence of the HNC/HCN ratio on the water production rate is seen.
Therefore we conclude that, as explained below, the apparent
distinction between the more productive and weak comets in
Figure~\ref{fig_prod} is simply caused by selection effects, as
all active comets in our sample happened to be observed at small
heliocentric distances.

\begin{figure}
\begin{center}
\epsfig{file=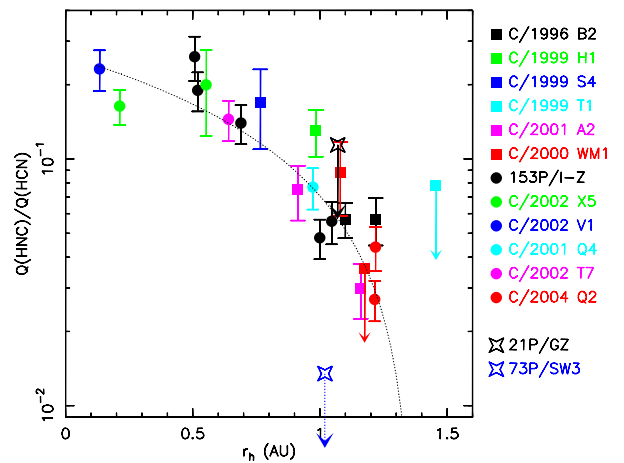,width=6cm,angle=270}
\end{center}
\caption{\label{fig_rh} Observed HNC/HCN abundance ratio in comets as
  a function of the heliocentric distance, $r_h$. The dotted line
  shows a linear fit to the data for $r_h$ between \about 0.5 and 1.2
  AU. The fitted slope is $0.192 \pm 0.021$; the observed variation of
  the HNC/HCN abundance ratio with $r_h$ is thus statistically
  significant.}
  \vspace{1em}
\end{figure}

Figure~\ref{fig_rh} shows the dependence of the HNC/HCN ratio
on the heliocentric distance of the comet. A clear variation is seen
and, and for $r_h$ between \about 0.5 and 1.2~AU, the ratio is well
described by the linear relation
\begin{equation}
Q_{\rm HNC}/Q_{\rm HCN} = 0.26 - 0.19 ~ r_h ({\rm AU})~,
\end{equation}
\noindent shown as the dotted line in Figure~\ref{fig_rh}. The
formal $1\sigma$ uncertainties of the intercept and the slope are
0.024 and 0.021, respectively. The two measurements at the smallest
heliocentric distances, $r_h$\about 0.1--0.2~AU (comets C/2002 X5 and
C/2002 V1, both with relatively high water production rates of \about
$4 \times 10^{30}$~mol\,s$^{-1}$), do fall slightly below the
best fit line. This may indicate that the dependence of the
HNC/HCN abundance ratio on the heliocentric distance flattens for $r_h
\la 0.5$~AU. The least squares fit given above is based on all
measurements in the 0.5--1.2~AU heliocentric distance range. However,
a similar $r_h$ dependence is obtained from measurements of comets
with the water production rates below $4 \times
10^{29}$~mol\,s$^{-1}$, which restricts the $r_h$ range to
0.75--1.2~AU (the slope given by the least squares fit is
$0.144\pm0.031$ vs. $0.191 \pm 0.021$ above; a \about 1.3$\sigma$
difference). The more productive comets in our sample thus
follow essentially the same HNC/HCN dependence on the heliocentric
distance, as derived for weak comets. The apparent distinction between
the two groups in Figure~\ref{fig_prod} is caused by the fact that all
more productive comets in our sample happened to be observed
preferentially close to the Sun.\footnote{Another way to separate
the dependence of the HNC/HCN abundance ratio on $r_h$ and $Q_{\rm
H_2O}$ is by fitting the data with a combined power-law in the form $R
= a r_h^b Q^c$, with $a$, $b$, and $c$ being free parameters. The
power-law fit diverges at small heliocentric distances, but in the
range of $r_h$ considered here, 0.5--1.2~AU, is a good representation
of the measurements. For $r_h$ in AU, and $Q$ in the units of
$10^{29}$~\ps, the best fit parameters are $a = 0.059 \pm 0.004$, $b =
-2.30\pm0.21$, and $c = -0.09 \pm 0.08$. This confirms the presence of
a strong dependence of the HNC/HCN ratio on the heliocentric distance
and no statistically significant trend with the water production
rate.} The observed variation in the HNC/HCN abundance ratio with the
heliocentric distance is likely related to the change in the
temperature of the grains in the coma, as the comet approaches the
Sun.

\begin{figure}
\begin{center}
\epsfig{file=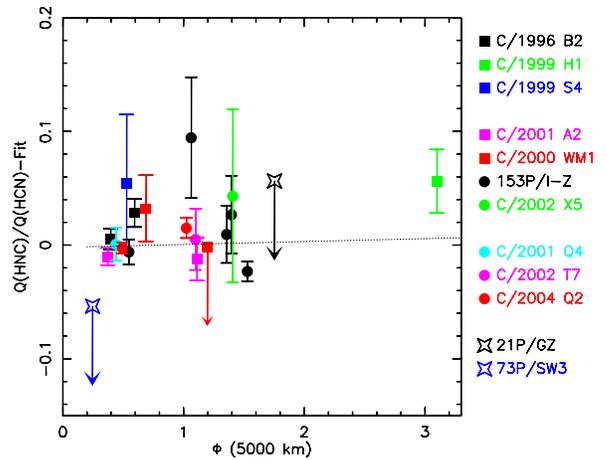,width=6cm,angle=270}
\end{center}
\caption{\label{fig_delta} Deviation of the HNC/HCN abundance ratio
  from the best fit line described in Fig.~4, in comets with $r_h$
  between 0.5 and 1.2~AU, plotted as a function of the telescope field
  of view (in the units of 5000 km). The dotted line shows a formal
  least-squares fit to the data. The fitted slope is $0.0025 \pm
  0.0063$; no statistically significant trend is seen.}
\end{figure}

The 3-$\sigma$ upper limit of 1.1\% for the HNC/HCN ratio in comet
73P/Schwassmann-Wachmann based on our observations is \about 7 times
lower than the typical value at 1~AU given by the least-squares fit to
the measurements in other comets (dotted line in Fig.~\ref{fig_rh}).
Given the small geocentric distance of comet 73P at the time of our
observations (0.072~AU), the HNC production rate could be
significantly underestimated if the emission originates from an
extended source in the coma. The production rates given above are
computed assuming a parent density distribution for both molecules
(direct release from the nucleus). The HNC production rate would
increase by a factor of 7 if a daughter density distribution with a
parent scale-length of \about 5000~km is assumed. The resulting
HNC/HCN abundance ratio in comet 73P would then be consistent with
other measurements in moderately active long-period comets at 1~AU.
Observations of comet Hale-Bopp with the Plateau de Bure
Interferometer imply an HNC parent scale-length $\ll 2000$~km
\citep{bockel05, bockel07}. The HNC parent scale-length has never been
constrained directly by interferometric observations in moderately
productive comets. However, observations of comet 153P/Ikeya-Zhang
presented by Biver et al. (2006b) suggest that the HNC parent
scale-length in this comet was less than 2500~km (equivalent to less
then 7500~km when scaled to 1~AU), with the best fit $L_p = 0$. To
further investigate the effects of a possible extended source of HNC
on the measured HNC/HCN abundance ratios, we plot in
Figure~\ref{fig_delta} the HNC/HCN ratio in the long-period comets
observed to date as a function of the telescope field of view (the
linear trend with the heliocentric distance given by eq. (1) has been
subtracted out). The formal linear fit is shown as a dotted line; no
statistically significant trend is seen. A combined power-law fit to
the HNC/HCN abundance ratio, as a function of the heliocentric
distance and the field of view, $\Phi$ (in the units of 5000 km), in
the form $R = a r_h^b \Phi^c$, gives $a=0.058 \pm 0.004$, $b = -2.03
\pm 0.21$, and $c = 0.10 \pm 0.12$. This confirms that no trend with
the field of view is apparent in the data. We note that other comets
in our sample, including comet Hyakutake, were observed with a
telescope field of view \about 2000~km. If the HNC parent scale-length
at 1~AU is indeed of order 5000~km, the HNC production rates and the
resulting HNC/HCN abundance ratios in these objects would scale up
accordingly. While only future sensitive interferometric observations
will be able to unequivocally constrain the HNC parent scale-length, a
possible extended source of HNC cannot explain the discrepancy between
the HNC/HCN ratio in comet 73P and the values measured in other
moderately active comets at 1~AU.

Comet 73P/Schwassmann-Wachmann is the first short-period
(Jupiter-family) comet in which a sensitive upper limit for the
HNC/HCN ratio has been obtained. Our observations might thus indicate
that HNC is depleted in Jupiter family comets, for example due to
evolutionary effects in the ices in the outer layers of the nucleus.
However, comet 73P was in the process of break-up and fragmentation,
with fresh unprocessed material exposed for the first time to solar
radiation. Infrared and radio observations \citep{dello07, biver06a}
show remarkable similarity in the composition of fragments B and C.
All measured volatiles, except HCN, were depleted with respect to
water in both fragments, compared to other comets. Furthermore, no
evidence of temporal variations in the compositions of the fragments
during and in-between outbursts was seen. The explanation of the low
HNC/HCN ratio in comet 73P/Schwassmann-Wachmann as due to evolutionary
effects in the surface layers of the nucleus thus appears unlikely.

A comparison of the HNC abundance in comet 73P/Schwassmann-Wachmann
with that measured in comet C/1999 S4 (LINEAR) is particularly
interesting, as both comets were in the process of break-up, with
fresh material exposed to the solar radiation. Based on their IR
measurements, \cite{dello07} argue that relative abundances of
parent volatiles in comet 73P/Schwassmann-Wachmann (with the
exception of HCN) resemble those in C/1999 S4 and are in the depleted
range with respect to the majority of comets characterized to
date (see also \citealt{villa06}). HNC was detected in C/1999 S4 with
an abundance of \about 17\% with respect to HCN at a heliocentric
distance of 0.77~AU in a small field of view of 2600~km
\citep{bockel01}. Correcting for the difference in the heliocentric
distance, as given by eq. (1), this is a factor of \about 11 higher
than the 3-$\sigma$ upper limit in 73P/Schwassmann-Wachmann. The HNC
abundance relative to water is also higher in C/1999 S4, by a factor
of \about 8--9; $1.7 \times 10^{-4}$ \citep{bockel01} vs. $\la
2\times 10^{-5}$ (Biver et al. in prep.) We next consider coma
  chemistry versus gas-grain processes as means of producing HNC.

\section{Origin of Cometary HNC}

\cite{rodgers01b} considered a wide range of possible explanations for
the origin of HNC in moderately active comets. They ruled out the
production by ion-molecule chemistry or by reactions of energetic
hydrogen atoms with HCN. They also ruled out photo-dissociation of
small organic molecules, such as CH$_2$NH, HNCO, and other N-bearing
compounds. This left large organic molecules or polymers as possible
parents of HNC. One possible candidate was hexamethylenetetramine
(HMT; C$_6$H$_{12}$N$_4$), produced by UV irradiation of simple ices
containing ammonia, water, formaldehyde and methanol. However this
compound was shown to be stable to thermal degradation \citep{fray05}
and therefore an unlikely parent of HNC. \cite{rodgers01b} also
suggested HCN polymers (or oligomers), produced from irradiated HCN
molecules in the nuclear or pre-cometary ices, as a possible parent of
HNC (once HCN tetrameters are formed, they can easily polymerize to
polyaminocyanomethylene (PACM; [-(NH$_2$)C(CN)-]$_{\rm n}$), which can
then undergo ring closure to form ladder polymers). If such polymers
are present in cometary nuclei, one would expect their abundance in
the coma to be correlated with that of HCN. The abundance of HCN in
comet 73P/Schwassmann-Wachmann implied by IR and radio observations
\citep{biver06a, villa06, dello07} is near the high end of
measurements in \about 30 comets observed at radio wavelengths
\citep{biver02, biver06b}. PACM polymers should thus be quite abundant
in this particular comet and the low HNC abundance implied by our
measurements appears to argue against PACM as a source of cometary
HNC.

A different class of HCN polymers, formed not from HCN molecules, but
instead from ammonia and acetylene, has been invoked to explain the
low $^{14}$N/$^{15}$N ratio observed in CN in several comets, a factor
of two below that measured in HCN in comet Hale-Bopp (e.g.,
\citealt{arpig02}). In this scenario, some CN is produced from
photo-dissociation of HCN, however a secondary, presumably
polymeric, source of CN is needed. In cold, dense interstellar
material, large $^{15}$N enhancements may be present in ammonia ices
\citep{charnley02b}. These ices could be precursors of cometary
material. Further energetic processing is required to incorporate this
nitrogen reservoir into organic material (e.g., addition of NH$_2$
side-groups to PAH molecules; \citealt{charnley02c}). The
secondary source of CN in cometary comae may thus be a polymer
produced from ammonia, since only interstellar ammonia can obtain
sufficiently low $^{14}$N/$^{15}$N ratios, and from carbon compounds,
such as acetylene, methane, or ethane.

The initial analysis of the Stardust organic data \citep{sandford06}
indicates that a significant fraction of organic nitrogen is in a form
of aliphatic carbon, with methylamine (CH$_3$NH$_2$) and ethylamine
(CH$_3$CH$_2$NH$_2$) being the two major carriers identified, with
similar abundances. These species are very volatile and
\cite{sandford06} argue that they are likely present in the cometary
material in a form of an amine-rich organic polymer rather than as a
free primary amine. One may speculate that energetic processing of
ices containing ammonia, methane, and ethane could produce these
molecules, as well as perhaps some organic polymers. Laboratory
studies are urgently needed to test this idea. If such polymers are
also a source of cometary HNC, the low HNC abundance in comet
73P/Schwassmann-Wachmann appears consistent with the observed
depletion of ammonia and acetylene implied by IR observations
\citep{dello07}.
We note that the rotational spectrum of
methylamine in the 49--326~GHz frequency range has now been measured
in the laboratory \citep{ilyushin05}. It is thus feasible to search
for this species in relatively active comets.

A comparison of the HNC/HCN ratio in comet 73P/Schwassmann-Wachmann
with measurements in other members of the 21P/Giacobini-Zinner
compositional class is of great interest. This class includes objects
with extreme C$_2$ and NH$_2$ depletions that are related to
C$_2$H$_2$ and NH$_3$ depletions in nuclear ices. We would expect a
low HNC/HCN ratio in these objects as well. So far, very few members
of this class have been identified \citep{fink96}. The upper limit for
the HNC/HCN ratio in Giacobini-Zinner obtained during its 1998
apparition ($<$ 11\%, \citealt{biver99}; Fig.~3) is not sensitive
enough to draw definitive conclusions. Comet 21P/Giacobini-Zinner will
make a favorable apparition in 2018 (at 0.4~AU from the Earth), during
which sensitive measurements will be possible using the Atacama Large
Millimeter/Submillimeter Array (ALMA).

\section{Summary}

Observations of hydrogen isocyanide in cometary atmospheres carried
out to date indicate that the HNC production has to be efficient in
the inner coma, just as the material leaves the nucleus. The process
has to be temperature dependent to explain the observed variation in
the HNC/HCN ratio with the heliocentric distance. Thermal degradation
of macromolecules or polymers produced from ammonia and carbon
compounds, such as acetylene, methane, or ethane appears consistent
with all existing data, including the very low HNC/HCN ratio in comet
73P/Schwassmann-Wachmann reported here. Such polymers have been
invoked previously to explain anomalous $^{14}$N/$^{15}$N ratios
measured in cometary CN.

Additional interferometric observations of HNC in comets are needed to
provide constraints on the spatial distribution of this molecule in
the cometary comae. Such observations will soon be possible in
moderately productive comets with (sub)millimeter interferometers
(e.g., e-SMA). Measurements of the H$^{15}$NC/H$^{14}$NC isotopic
ratio would be instrumental in determining whether the HCN polymers
similar to those invoked to explain the enhanced $^{15}$N abundance in
CN may also be a source of cometary HNC. However, such measurements
will have to await the commissioning of ALMA.

\acknowledgments This research has been supported by NSF grant
AST-0540882 to the Caltech Submillimeter Observatory. D.C.L.
acknowledges support from the Observatoire de Paris during his stay in
Meudon. S.B.C. acknowledges support from NASA's Planetary Atmospheres
Program. We thank N.~Fray for useful discussions and the referee, M.
DiSanti, for a thorough and helpful review of the manuscript.


\end{document}